\documentclass[aps,pra,twocolumn,10pt]{revtex4-2}
\usepackage{amsmath,amssymb,graphicx}
\usepackage[english]{babel}
\usepackage{braket}

\newcommand{\sub}[1]{_{\mbox{\scriptsize{#1}}}}

\begin{document}

\title{Coherent Control of Domain-Wall Transport in an Ultracold Bose Gas}
\author{O.~Farion}
\author{M.~Pourzand}
\author{J.M.~McGuirk}
\affiliation{Department of Physics, Simon Fraser University, Burnaby, British Columbia V5A 1S6, Canada}

\bibliographystyle{apsrev4-2}

\begin{abstract}
Domain walls are carriers of spin transport whose controlled manipulation underlies a wide range of spintronic and information-processing technologies. Here we demonstrate tunable domain-wall transport in a weakly interacting nondegenerate ultracold Bose gas. We initialize a three-domain pseudo-spin-1/2 texture and observe spontaneous propagation of long-lived domain walls driven by exchange-mediated spin currents. By varying the orientation of the spin domains, we control the balance of spin currents across the walls and thereby tune their trajectories, including reversals of the initial direction of motion. Measurements reveal a crossover from an exchange-stabilized regime, in which coherent spin-exchange collisions suppress wall motion, to a diffusion-dominated regime characterized by increasing velocities. Numerical solutions of a quantum Boltzmann equation reproduce the observed dynamics and identify transverse phase gradients as an important control parameter governing domain-wall propagation. These results establish coherence and phase engineering as tools for programming spin transport in ultracold gases and provide a route toward controllable domain-wall dynamics in atomtronic systems.
\end{abstract}

\maketitle

Control of domain-wall transport is a central challenge across systems ranging from magnetic materials to quantum fluids. In solid-state spintronic devices, domain walls serve as mobile information carriers whose trajectories are manipulated through magnetic fields \cite{OnoNanowire, Atkinson2003}, spin-transfer torques \cite{Berger1996SWemission, Yamaguchi2004driven, Hayashi2008}, and engineered spin textures \cite{Fert2017}. The ability to program domain-wall motion through internal spin degrees of freedom underlies proposed technologies such as racetrack memory \cite{Parkin2008racetrack, Parkin2015}, domain-wall logic \cite{AllwoodDWlogic}, and other emerging technologies. More generally, domain walls provide a model system for studying how microscopic spin transport gives rise to collective nonequilibrium dynamics \cite{Beach2008}.
	
Ultracold atomic gases offer a complementary platform for investigating domain-wall transport. Their interactions, geometry, and spin textures can be engineered with exceptional precision, enabling studies of collective spin dynamics that are difficult to realize in condensed-matter systems \cite{McGuirk2002, Levitov2002, Laloe2002, Thomas2008, Koschorreck2013, Sengstock2013, Trotzky2015, Chien2015}. In weakly interacting nondegenerate gases, coherent spin-exchange collisions generate collective spin currents through the identical spin rotation effect (ISRE) \cite{Bashkin1981, Lhuillier1982}, producing phenomena such as spin waves \cite{McGuirk2002, Levitov2002, Laloe2002}, metastable spin domains \cite{Lewandowski2002, Thomas2008}, and coherence-mediated spin transport \cite{Sommer2011, Niroomand2015}. Previous studies demonstrated that coherence within a domain wall can strongly modify diffusion and stabilize spin textures against decay \cite{Niroomand2015, Graham2018}. However, despite extensive work on collective spin transport, the controlled motion of domain walls themselves has remained largely unexplored.

Domain-wall dynamics have recently emerged as an important topic in quantum-degenerate gases, where superfluid hydrodynamics, phase coherence, and interaction-driven instabilities produce rich nonequilibrium behavior \cite{Chin2022,Cominotti2023}. These systems operate in a distinct transport regime, as opposed to the nondegenerate regime studied here, where thermal energies dominate all other relevant energies, and exchange-mediated spin currents generated by coherent collisions play a key role. Understanding whether domain-wall transport can be controlled in this regime remains an open question.
	
Here we demonstrate tunable domain-wall transport in a three-domain spin texture realized in an ultracold nondegenerate Bose gas. The gas is prepared in extended metastable regions of aligned spins (domains) separated by narrow regions of rapidly spatially varying spin (domain walls). We show that domain-wall trajectories are governed not only by population imbalances driving diffusive currents, but also by the coherence and phase geometry of the spin texture. By controlling the polarization of the central domain, we engineer exchange-driven spin-current imbalances that accelerate, reverse, or suppress domain-wall motion. Numerical solutions of a quantum Boltzmann equation further reveal that transverse phase gradients play an important role in determining wall trajectories, enabling the dynamics of one domain wall to be manipulated through the spin texture associated with another. The resulting motion exhibits a crossover between an exchange-stabilized regime, in which coherent spin interactions impede wall propagation, and a diffusion-dominated regime characterized by rapid transport.
	
These results establish domain walls as controllable carriers of spin transport in nondegenerate quantum gases and identify coherence and phase engineering as tools for programming collective spin dynamics. More broadly, they demonstrate a mechanism for controlling transport through internal quantum degrees of freedom, providing a new avenue for spin-based atomtronic architectures \cite{Amico2021}.

The experimental system consists of magnetically trapped $^{87}$Rb atoms in an axisymmetric quasi-one-dimensional hybrid Ioffe-Pritchard trap with frequencies $\{\omega_\rho, \omega_z\} = 2\pi \times \{250, 6.8\}$~Hz \cite{Graham2018}. The hyperfine ground states $\ket{1} \equiv \ket{F,m_f=1, -1} $ and $\ket{2}  \equiv \ket{2, 1} $ form a pseudo-spin-1/2 system coupled via a two-photon microwave transition. The spins evolve in a spin-independent potential derived from a mutual compensation scheme that balances differential mean-field and Zeeman shifts \cite{Lewandowski2002}. Atom clouds are laser and evaporatively cooled to peak density $n_0 = 2 \times 10^{13}$~cm$^{-3}$ and temperature $T=650$~nK, $\sim$~60\% above quantum degeneracy. Atoms in $\ket{1}$ are imaged destructively after adiabatic transfer to $\ket{2,-2}$, using the $\ket{2,-2}\rightarrow\ket{3,-3}$ cycling transition. $\ket{2}$ is measured separately by first inverting the population with a $\pi$-pulse prior to imaging. The longitudinal magnetization is then given by $M_\parallel(z,t) = N_2(z,t) - N_1(z,t)$, where $N_{j}$ (j=1,2) is the population of each state, normalized to the center of the ensemble. The transverse spin magnitude $M_\perp (z,t)$ and phase $\phi(z,t)$ are measured via Ramsey spectroscopy \cite{Graham2019thesis}.  

Spin textures are created by locally altering transition frequencies during a two-photon microwave $\pi$-pulse, using a digital micromirror device (DMD) and a 780~nm laser detuned halfway between the $F=1$ and $F=2$ D2 excited-state transitions. This spatially modulated light shifts the $\ket{1}-\ket{2}$ transition frequency via the AC Stark shift with a precisely chosen detuning of $\sqrt{3}\Omega\sub{R}$, where $\Omega\sub{R}$ is the resonant two-photon Rabi frequency. A resonant $\pi$-pulse transfers atoms in the dark regions to $\ket{2}$, while atoms in the bright regions experience an off-resonant $2\pi$-pulse and remain in state $\ket{1}$. A domain wall forms in regions where the optically-induced detuning varies from zero to $\sqrt{3}\Omega\sub{R}$. In the Bloch sphere picture, the magnetization vector wraps around the Bloch sphere from north to south poles in moving across the domain wall, with a width determined by the spatial change of the optical potential. To create high fidelity domains, we minimize intensity variations in the light pattern, e.g.~due to reflections from vacuum cell windows, by applying a correcting pattern to the DMD to produce homogeneous intensities across each domain \cite{Pourzand2022thesis}. Consisting of only rapid coherent rotations and negligible spontaneous emission, this domain initialization process preserves spin coherence, though ensuing atomic motion can lead to dephasing and decoherence.

The initial longitudinal magnetization of the three-domain spin texture is characterized with a phenomenological model
\small
\begin{equation}
  M_{\parallel}(z, 0)=M_0\tanh \Big(\dfrac{z-z_l}{\lambda_l}\Big) \tanh \Big(\dfrac{z-z_r}{\lambda_r} \Big) \exp \Big({\dfrac{-z^2}{2\sigma_z^2}}\Big) \label{eq:lgM},
\end{equation}
\normalsize
where $z_i$ and $\lambda_i$ are initial domain wall centers and widths respectively, $i=\{l,r\}$ subscripts indicate left and right walls, and $\sigma_z=190~\mu$m is the axial Gaussian cloud width. $z_i$ and $\lambda_i$ are controlled by patterns displayed on the DMD. In this work, $\lambda_i\approx 70~\mu$m; extremely small domain wall widths are avoided, as in a spin-independent potential they are subject to decay induced by transverse perturbations via the Castaing instability \cite{Castaing1984}.

Magnetization dynamics here is a competition between diffusion and exchange-driven currents. Diffusion equilibrates inhomogeneous spin textures, while coherent exchange-mediated dynamics can stabilize them. The ISRE causes spin currents to precess about the local magnetization \cite{Williams2002}, which can stabilize a spin texture against diffusion when the magnetization varies sufficiently slowly in space. Adding spatially varying effective magnetic fields can augment this effect by rotating transverse spin components to better match the phase winding of the domain wall magnetization, or vice versa  \cite{Graham2018}. If the system has sufficient coherence, collective behavior tends to dominate diffusive relaxation. The ISRE strength is characterized by the Leggett-Rice parameter, $\mu=\omega\sub{ex}\tau$, which is the ratio between exchange collision ($\omega\sub{ex}$) and elastic collision ($1/\tau$) rates. Here the peak $\mu\simeq12$, indicating the strong influence of exchange collisions on the dynamics.

\begin{figure}[!htb]
\includegraphics[width=0.47 \textwidth]{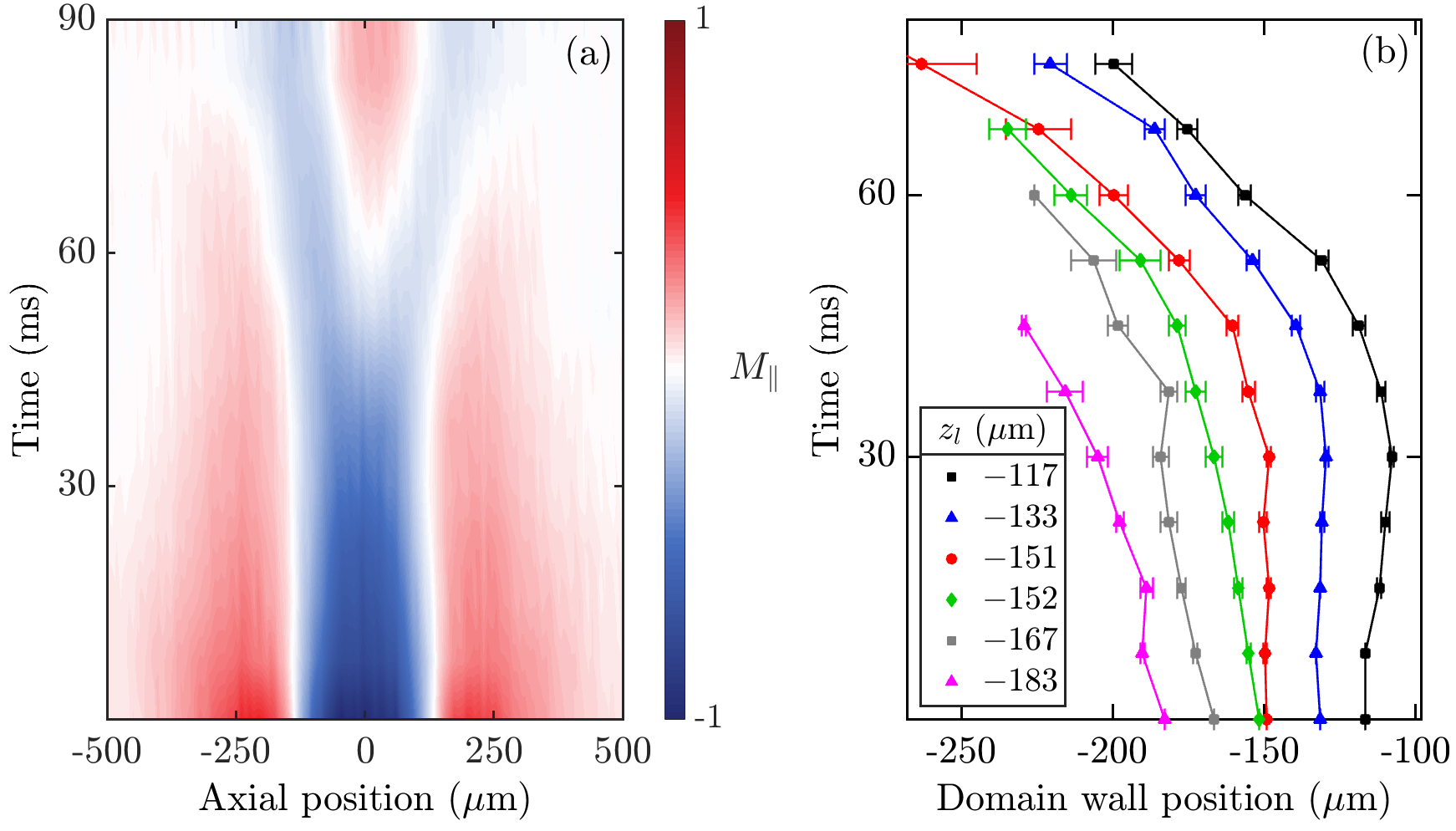}  
\caption{ \label{fig:fig1} (a) Spatiotemporal evolution of radially averaged longitudinal magnetization $M_\parallel(z,t)=N_2-N_1$ exhibiting domain wall movement. Data is averaged over 3 measurements at each time and interpolated between time steps $\delta t=7.5$~ms. (b) Domain wall trajectories extracted from $M_\parallel(z,t)$ using a threshold method, for a range of initial domain wall positions $z_l$.} 
\end{figure}

An example of the spatiotemporal evolution of $M_\parallel (z,t)$ is shown in Fig.~\ref{fig:fig1}(a). As the spin distribution evolves, the phenomenological form of Eq.~\eqref{eq:lgM} no longer describes $M_\parallel$. Instead, we extract the wall positions using a threshold algorithm with a threshold value $M_\parallel^{\mbox{\scriptsize{thr}}}$ and identify regions where $\lvert M_\parallel \rvert < M_\parallel^{\mbox{\scriptsize{thr}}}$. The centers of these regions are taken to be the domain wall centers. The threshold is chosen to identify the zero crossing ($M_\parallel=0$) robustly despite noise, and the half-width of the selected region is used as the positional uncertainty. Though spin textures are initialized in domains that are nominally symmetric around the trap center, due to slight asymmetries analysis is limited to domain walls with negative initial positions (``left''). The effect of these asymmetries on the evolution of $M_\parallel$ is discussed later. In some realizations, an additional domain forms near the trap center at later times [Fig.~\ref{fig:fig1}(a)]. Its origin is not yet understood; we do not observe this behavior in homogeneously polarized samples, and simulations indicate the domain-wall transport trends discussed below are robust to its presence.

Examples of domain wall trajectories are shown in Fig.~\ref{fig:fig1}(b). The initial domain-wall position determines the size of the domains and thus the overall spin balance in the system. When the initial population is significantly imbalanced between the two states, transport driven by the diffusive motion of atoms dominates and domain walls drift rapidly from their initial position with a direction determined by the population imbalance, as seen in the left-most trajectories. More complex spin-current dynamics arise when spin textures have nearly equal initial spin population ($N_1\approx N_2$), corresponding to $|z_i| \approx 135\ \mu$m. In this regime, domain walls typically move slowly at first, before accelerating at later times. This behavior is in line with earlier observations that the presence of a significant amount of coherent exchange collisions in the domain wall dramatically slows spin dynamics \cite{Niroomand2015}. In this regime, the character of this motion is largely independent of the exact initial wall position, and thus the precise global spin balance as well. In the early, slow phase of motion, some walls are even seen to change the direction of motion (e.g.~at $\sim$40~ms), which demonstrates that the direction of spin currents through the wall can spontaneously reverse. To understand this type of motion, we investigate the role of the spin domain orientation.

\begin{figure}[!htb]
\includegraphics[width=0.48\textwidth]{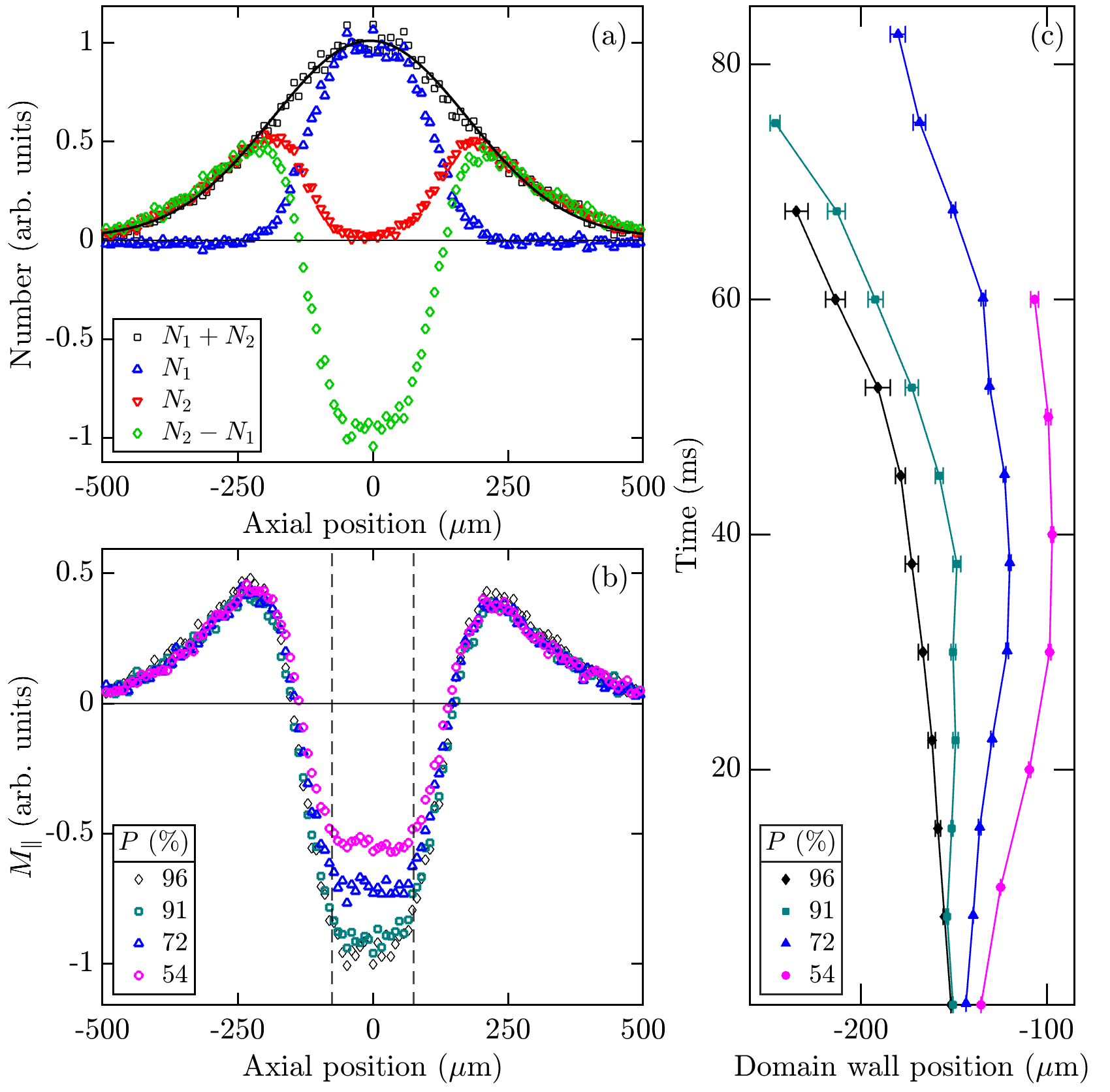} 
 \caption{\label{fig:fig2} (a) Initial populations of state $\ket{1}$ and $\ket{2}$. The sum is a Gaussian given by the harmonic trap parameters, and their difference is $M_\parallel$. (b) Initial $M_\parallel(z)$ profiles prepared with various rotated middle domain orientations. Dashed vertical lines indicate the region in which $P$ is calculated, $z_l + \lambda_l/2<z< z_r - \lambda_r/2$. (c) Domain wall trajectories corresponding to domain orientations shown in (b). Rotated domains (smaller $P$) lead to larger spin current imbalances, increasing inward motion at short time.}
\end{figure}

As described above, the initialization sequence is coherent and leaves domains fully polarized. However their orientation may not be perfectly longitudinal, either by design or due to small imperfections in the optical potentials used for initialization. In particular, the optical potential is made uniform along the axial direction during initialization using an iterative method, but inhomogeneities in beam intensity along the radial axis are not imaged, thus limiting the ability to form perfectly orthogonal longitudinal domains. This effect manifests in the center domain, where the optical potential is present. The amount of domain rotation in the center domain is characterized by the normalized longitudinal projection of the magnetization, $P=M_\parallel/(N_1+N_2)$, in the middle region of the center domain, shown in Fig.~\ref{fig:fig2}. Additionally, the orientation of the middle domain can also be intentionally altered by varying the AC Stark shift during the preparation via the overall light intensity. Examples of initial preparations with intentionally rotated polarizations are shown in Fig.~\ref{fig:fig2}(b). Corresponding domain wall trajectories are shown in Fig.~\ref{fig:fig2}(c), demonstrating that the change in polarization rotation significantly impacts the direction and the rate of spin currents through the wall. This behavior can be understood by noting that when $P$ deviates significantly from 1, the magnetization vector has a significant transverse component. Therefore, exchange collisions are enhanced within the domain wall, leading to the conversion of transverse magnetization into longitudinal and vice versa, which increases the total spin currents across the domain wall. Thus, decreasing $P$ in the center domain drives the domain wall towards that region initially.

\begin{figure}[!htb]
\includegraphics[width=0.5\textwidth]{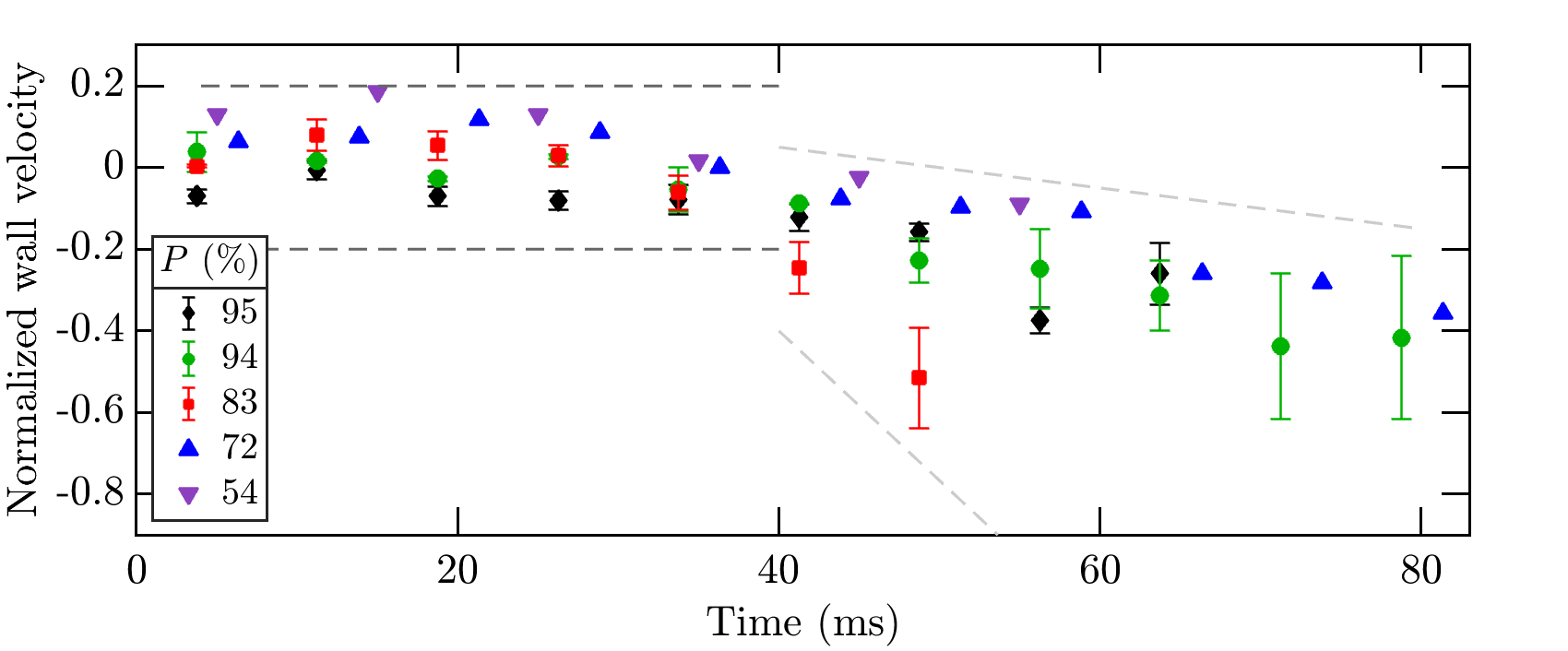} 
       \caption{\label{fig:fig3} Domain wall velocity dynamics for various initial conditions, normalized by the thermal velocity $v\sub{t}$. Points with error bars are averaged over 2-4 trials, with error bars indicating the standard error of the mean. Points without error bars represent a single trial. Dotted lines guide the eye to distinguish between two regimes of motion. At short times, exchange collisions stabilize the domain wall and suppress its motion. At later times, diffusion becomes dominant and the wall velocity increases. The time scales of the regimes can be tuned via the initial coherence in the domain wall.}
 \end{figure}

Regimes of domain-wall motion are identified by examining the wall velocity $v_{\mathrm{dw}} = \delta z_l/\delta t$. Fig.~\ref{fig:fig3} shows the evolution of $v\sub{dw}$ normalized by the mean thermal velocity $v\sub{t}=\sqrt{16 k_\mathrm{B} T/(3 \pi m\sub{Rb})}$, with atomic mass $m\sub{Rb}$. The velocity dynamics display two qualitative regimes of domain-wall motion, consistent across a range of initial wall positions and domain polarizations. The first regime is characterized by both small velocity and acceleration due to exchange collisions in the domain wall providing a spin-stiffness that stabilizes domains against diffusion. At later times, diffusion drives increasingly rapid wall motion. Near the low-density edge of the cloud, collision rates become small and motion crosses toward the collisionless, ballistic regime, with the wall velocity approaching the characteristic thermal speed.

The timescales of these regimes are largely determined by the initial coherence in the domain wall. Earlier work showed that greater coherence in the domain wall leads to longer time scales for longitudinal diffusion, due to transverse-spin mediation of longitudinal diffusion (see \cite{Niroomand2015}). Complementary measurements of the spatiotemporal evolution of transverse magnetization reveal a significant decrease of the coherence in the domain walls at $\sim40$~ms due to a collapse in coherence caused by counterpropagating orthogonal spin currents (see \cite{McGuirk2010collapse}). This drop in coherence reduces the local mean field in the domain wall and signifies the end of the first stage. Differences in acceleration during the second stage can be explained by varying levels of initial coherence and ensuing decoherence rates, as well as the importance of the transverse phase gradient in the domain wall, which is discussed below.

The velocity dynamics shown in Fig.~\ref{fig:fig3} demonstrate that rotated domain polarizations typically lead to larger domain wall velocities, driven by increased imbalance of the spin currents across the wall. This behavior has strong dependence on both domain-wall coherence and domain polarization rotation. Lower coherence, for example, leads to faster motion approaching classical diffusion time scales. For highly rotated polarizations, the initial transverse magnetization magnitude in the middle domain is large, which leads to qualitatively different longitudinal spin dynamics, such as the pronounced inward motion seen in Fig.~\ref{fig:fig2}(c). Consequently, the velocities from different initializations $P$ exhibit more spread at longer times. 

\begin{figure}[!htb]
\includegraphics[width=0.5\textwidth]{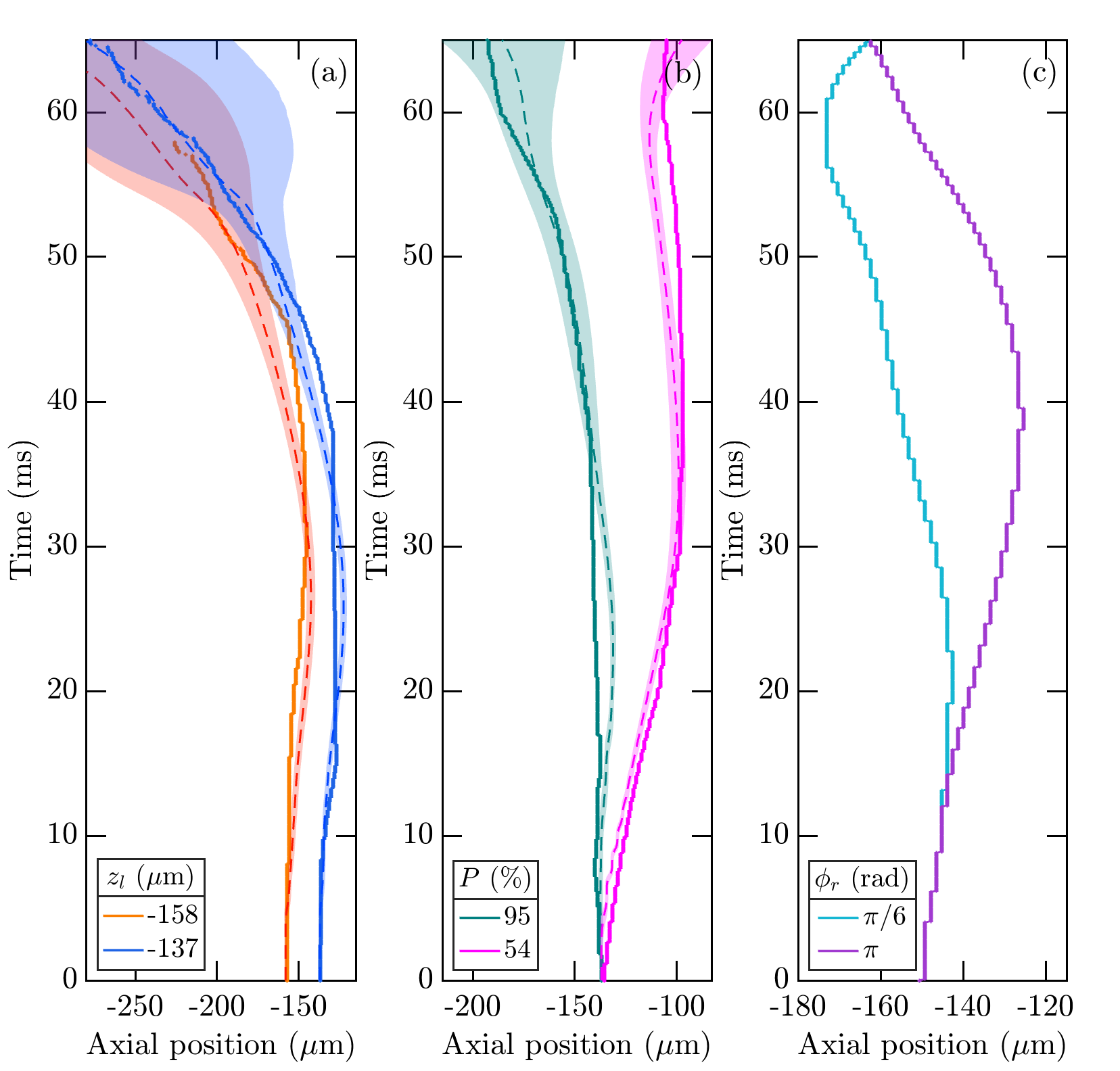} 
       \caption{Comparison of typical measured trajectories (solid lines) and simulated trajectories (dashed lines), using experimentally determined initial parameters. Inward (outward) wall motion is to the right (left). Shaded regions show simulation uncertainty. Simulations confirm that (a) domain wall motion is similar for a range of initial wall positions, but (b) that altering domain polarization leads to qualitatively distinct dynamics. (c) Comparison of simulated left domain wall trajectories with identical initializations, except for different transverse spin textures initialized in the right wall. \label{fig:fig4} }
 \end{figure}

To verify the experimental study of the dependence on the initial parameters of longitudinal magnetization and to further examine the role of initial transverse magnetization, we also perform numerical simulations of the domain wall dynamics (Fig.~\ref{fig:fig4}). The magnetization density distribution $\vec{m}(p,z,t)$ in an ultracold weakly interacting gas is described by a 1D quantum Boltzmann equation \cite{Levitov2002,Laloe2002,Williams2002}
\begin{equation}
  \partial_t \vec{m} +  \partial_0 \vec{m} - \dfrac{1}{\hbar} g \vec{M} \times \vec{m} = \partial_t \vec{m} |\sub{coll} \label{eq:Boltz},
\end{equation}
where  $\partial_0 = \dfrac {p}{m\sub{Rb}} \partial_z - m\sub{Rb} \omega_z^2 z \partial_p$ and $\partial_t \vec{m} |\sub{coll}$ represents collisional relaxation, modeled as the difference between the magnetization density and the local magnetization average, scaled by the local elastic collision time ($\sim4$~ms peak here) \cite{Williams2002}. The mean-field coupling constant is $g=8\pi\hbar^2a_{12}/m\sub{Rb}$, for $s$-wave scattering length $a_{12}$. The kinetic equation is solved numerically to find the magnetization vector, $\vec{M}(z,t) = \int dp\,\vec{m} (p,z,t)/ 2 \pi \hbar$, and extract the domain wall trajectories. The initial longitudinal magnetization follows Eq.~\ref{eq:lgM}, and the initial transverse magnetization is modeled by $M_\perp(z) e^{i\theta(z)}$, where $M_\perp(z)$ incorporates domain polarization $P$, as well as dephasing and decoherence during and immediately following domain preparation. The transverse magnetization orientation $\theta(z)$ is approximated by fixing the phase of the middle domain and setting the relative orientation of each side domain ($\phi_{l/r}$) and the phase gradients across each domain wall ($\partial_z\phi_{l/r}$). 

To compare simulation and experiment we use initial conditions taken from fits to the initial domain preparations. Longitudinal parameters $\{z_l, z_r, \lambda_l, \lambda_r, P\}$ are obtained from fits of the measured initial longitudinal magnetization. The initial transverse magnetization orientation is guided by Ramsey spectroscopy; however, it is one of the dominant sources of uncertainty due to its sensitivity to the exact amount of light used in domain preparation as well as measurement challenges in regions where $M_\perp$ is small. 

Numerical simulations for different initial domain wall positions are compared to experimental data in Fig.~\ref{fig:fig4}(a). Simulations confirm that the transport dynamics is similar for domain walls near $|z_l| \approx 0.8\ \sigma_z$, where $N_1 \approx N_2$. Fig.~\ref{fig:fig4}(b) shows comparisons of domain walls dynamics with different domain orientations $P$. Here also there is both qualitative and quantitative agreement -- faster wall propagation at short times occurs when the center domain is rotated to give more transverse spin, thus enhancing exchange-mediated transport.

The uncertainties in numerical solutions shown in Fig.~\ref{fig:fig4} are calculated from Monte Carlo simulations that include uncertainties from measured transverse phase gradients $\partial_z \phi_{l/r}$ and statistical and systematic fluctuations of temperature and density. $10^4$ trajectories are generated to find mean domain wall positions and variances. Each trajectory is terminated if, at any time, the wall is ejected from the cloud or a cloud-wide domain oscillation occurs. This procedure slightly biases the mean trajectory towards trajectories with longer lifetimes.

Due to experimental challenges in precisely controlling transverse magnetization in the domain wall, simulations are used to explore its role. Of particular interest is the role of asymmetries in transverse spin between domain walls, as seen in the asymmetry in the evolution of longitudinal magnetization in Fig.~\ref{fig:fig1}(a). Fig.~\ref{fig:fig4}(c) shows an example simulation of two identically prepared left domain walls, where the right domain wall was initialized with different transverse spin orientations $\phi_r$, with disparate trajectories. Exchange-mediated spin currents develop after a short initial period and drive different dynamics. Furthermore, varying $\partial_z \phi_{l/r}$ in simulations reveals the initial domain wall velocity is influenced not only by the relative domain orientation, but also by the magnitude of the phase gradients in the domain wall. Increasing the phase gradient can impede the adiabatic rotation of spin as atoms move from one domain to another, which enhances dephasing and speeds wall motion \cite{Graham2018}. When the phase gradients across the two domain walls differ, the wall with the larger initial gradient exhibits faster spin transport and dominates the global behavior of the spin ensemble. Thus, the dynamics of one domain wall can be tuned by controlling the phase texture in the second wall. 

Another potential source of disagreement between simulations and experiment is the domain wall coherence. Despite a fully coherent initialization sequence, measurements of the initial coherence in the walls range from $60\%-90\%$, due to both inhomogeneities in the optical potential and motion during initialization. Additional discrepancies may be introduced by further thermal motion averaging spin textures during the cloud expansion required for imaging, affecting the coherence inferred from the images. Furthermore, loss from dipolar relaxation and three-body collisions is neglected in the model.

We have demonstrated that domain-wall transport in a nondegenerate ultracold Bose gas can be controlled through the coherence and phase structure of a spin texture. By varying the polarization of the domains, we engineer spin-current imbalances that reverse, suppress, or accelerate wall motion, while numerical simulations identify transverse phase gradients as a nonlocal control parameter through which one domain wall influences the transport dynamics of another. The observed crossover from an exchange-stabilized regime to diffusion-dominated motion reveals how coherent spin-exchange interactions regulate collective transport far from equilibrium. More generally, these results establish domain walls as programmable carriers of spin transport in ultracold gases and demonstrate how coherence and phase engineering can be used to control collective nonequilibrium dynamics. This work identifies a new mechanism for manipulating transport in interacting quantum systems and provides a foundation for future studies of coherence-controlled spin transport and atomtronic architectures.

\emph{Acknowledgments.}
O.F. and M.P. contributed equally to this work. This research was supported by NSERC.

\emph{Data availability.}
The data that support the findings of this article are not publicly available but are available from the authors upon reasonable request.

\bibliography{References}
\end{document}